\documentclass[twocolumn,preprintnumbers,prl,amssymb]{revtex4}

\usepackage[dvips]{graphicx}
\usepackage{amsthm}
\usepackage{amssymb}
\usepackage{bm}
\usepackage{amstext}
\usepackage{psfrag}
\usepackage{amsmath}
\usepackage{amsfonts}
\newcommand{\p}{\partial}

\newcommand{\dd}{{\rm d}}

\newtheorem{theorem}{Theorem}[section]

\newtheorem{lemma}[theorem]{Lemma}

\begin{document}
\title{A topological interpretation of the color charge\footnote{Published at: Mod. {P}hys. {L}ett. {A} 22 (2007)
1143-1151}}

\author{E. Minguzzi\footnote{New permanent address: Departimento di Matematica Applicata, Universit\`a degli Studi di Firenze,
Via S. Marta 3, I-50139 Firenze, Italy}}
 \affiliation{Departamento de Matem\'aticas,
Universidad de Salamanca, \\ Plaza de la Merced 1-4, E-37008 -
Salamanca, Spain \\ and INFN, Piazza dei Caprettari 70, I-00186
Roma, Italy
\\ minguzzi@usal.es }

\author{C. Tejero Prieto}
 \affiliation{Departamento de Matem\'aticas,
Universidad de Salamanca, \\ Plaza de la Merced 1-4, E-37008
Salamanca, Spain
\\ carlost@usal.es }


\begin{abstract}
\noindent

We develop a theory  on a topologically non-trivial manifold which
leads to different vacuum backgrounds  at the field level. The
different colors of the same quark flavor live in different
backgrounds generated by the action of the torsion subgroup of
$H^{2}(M,2\pi\mathbb{Z})$ on $H^{2}(M,2\pi\mathbb{Z})$ itself. This
topological separation leads to a quark confinement mechanism which
does not apply to the baryons as they  turn out to live on the same
vacuum state. The theory makes some topological assumptions on the
spacetime manifold which are compared with the available data on the
topology of the Universe.

\end{abstract}


 \maketitle

{\small \noindent {\bf Mathematics Subject Classification  (2000)}.
57R15, 57R22, 57R56.

\noindent {\bf Key Words}. Flat bundles, torsion subgroup, quark
confinement, charge quantization. }

\section{introduction}

The Standard Model is based on the gauge group $U(1) \otimes
SU(2)\otimes SU(3) $. In this work we are going to exploit the
relation between the $U(1)$ and the $SU(3)$ sectors in spacetime
manifolds $M$ with non-trivial topology. The idea, clarified below,
is that each particle can be identified, at the electromagnetic
level, with a cohomology class on $H^{2}(M,2\pi\mathbb{Z})$.  If the
topology is not trivial and $H^{2}(M,2\pi\mathbb{Z})$ admits a
torsion subgroup of order three then there is a natural action of
this subgroup on $H^{2}(M,2\pi\mathbb{Z})$ which produces orbits
with three elements. The three cohomology classes identify three
`particles' that behave in the same way but that do not coincide. We
identify these three particles with the three colors of the same
quark flavor. Thus for every cohomology class on
$H^{2}(M,2\pi\mathbb{Z})$ we have three equivalent copies which are
the different colors of the `same' particle. The non-triviality of
the torsion subgroup implies that the different color fields are not
described by the usual electromagnetic gauge theory unless the gauge
group is enlarged to include a $SU(3)$ factor. In this case the
usual gauge theory is restored although it turns out to be based on
a $SU(3)$ principal bundle which is non-trivial. Far from being a
drawback this non-triviality is at the hearth of a possible
confinement mechanisms which we outline in section \ref{loj}.

The close relationship obtained here between the (topological)
aspects involved in the  $U(1)$ sector and  the $SU(3)$ sector was
expected to be  needed \cite{thooft76,lazarides82}  in order to
reach a deeper understanding of  electric charge quantization and
quark confinement.

%
%
%
%

Remarkably, the developed theory  has cosmological consequences as
it is based on some hypothesis on the topological structure of
spacetime. The connection with cosmology is clear under the
assumption that the non-trivial topology manifests itself at the
largest length scales. Under this assumption our particle model
could have been falsified by cosmological observations, and provides
an example of interaction between very small scale physics and very
large scale physics. These aspects are considered in the last part
of the work where it is shown that the latest cosmological data are
compatible with our model.

\section{Particles and cohomology classes}
 For simplicity in what
follows we shall ignore the $SU(2)$ sector of the Standard Model.
Alternatively the reader can think of the following fields as the
right-handed components of the particle considered, since these
components in the Standard Model transform with the $SU(2)$ trivial
representation. As a consequence of our simplification the $U(1)$
sector can be identified with the electromagnetic sector.

In order to fix the ideas let us focus on the quark flavor $\bar{d}$
of charge $q=e/3$, and on its three colored fields $\bar{d}(r)$,
$\bar{d}(g)$ and $\bar{d}(b)$. According to the usual treatment of
the $U(1)$ gauge theory \cite{kobayashi63} the matter fields are
sections of a suitable 1-dimensional vector bundle  associated with
a universal $U(1)$ bundle $Q$. On the bundle $Q$ a (electromagnetic)
connection 1-form $\omega$ can be defined and all the fields can be
considered as suitable sections of vector bundles associated to it
through a representation $\rho_{n} : U(1) \to \mathbb{C}$ of the
form $u \to u^n$ where $nq$, $n \in \mathbb{N}$, is the (electric)
charge of the particle. Sometimes it can be convenient to regard a
particle as a section of a vector bundle associated to the bundle
$Q^{n}$ under the trivial representation $\rho_1$. We shall use this
point of view as it allows to associate each kind of particle with a
certain principal $U(1)$ bundle. For instance, the three colored
components of $\bar{d}$ live on the same vector bundle. We denote
this fact by writing down the $U(1)$ bundles associated to them
\begin{equation}
\bar{d}: \quad Q,\quad Q, \quad Q.
\end{equation}
For the antiparticle  $d$ and the flavor $u$ we have
\begin{eqnarray}
d&:& \quad Q^{-1},\quad Q^{-1}, \quad Q^{-1},\\
u&:& \quad Q^{2}  \ \ ,\quad Q^{2} \ \  , \quad Q^{2}.
\end{eqnarray}
The manifold $M$  admits  a good covering $\{ U_i\}$. On each open
set $U_i$ there is a local potential 1-form $A^{i}$ such that on the
intersections $U_{ij}=U_{i} \cap U_{j}$, $\dd A^{i}=\dd A^{j}$, that
is the curvature 2-form of the $U(1)$ sector is globally well
defined. These 1-forms are the pullbacks of the connection 1-form
$\omega$ under the trivializing sections $\sigma_i: U_i \to Q$. Let
$g_{ij}= \sigma_i\circ \sigma_j^{-1}=e^{i\beta_{ij}}$ be the
transition functions. A particle $\psi$ is represented on each set
$U_i$ by a field $\psi^{i}$ that in the intersection $U_{ij}$
transforms as \footnote{We have removed the fundamental charge $q$
from this transformation by absorbing it into the definition of
gauge potential, $q A \to A$.}
\begin{eqnarray}
A^{i}&=&A^{j}+\dd \beta^{i j} , \\
\psi^{i}&=& e^{i n \beta^{i j}} \psi^{j} .
\end{eqnarray}
In particular the colored components associated with the same quark
flavor, say $u$, transform in the same way
\begin{eqnarray}
u^{i}(r) &=& e^{i2\beta^{i j}}  u^{j}(r), \label{q1} \\
u^{i}(g) &=& e^{i2\beta^{i j}}  u^{j}(g) , \label{q2}\\
u^{i}(b) &=& e^{i2\beta^{i j}}  u^{j}(b). \label{q3}
\end{eqnarray}
However, this formalism although successful   does no accomodate in
a simple way the observational fact that there are  actually two
fundamental electric charges, say the fundamental unit $q$, and the
fundamental unit for baryons and leptons $e=3q$.

Consider the short exact sequence
\begin{equation}
0\to {2\pi}\mathbb{Z}\to \mathbb{R} \xrightarrow{\exp(ix)} U(1) \to
1 ,
\end{equation}
where we  identify $U(1)$ and $\mathbb{R}/{2\pi}\mathbb{Z}$. It
gives rise to the long exact sequence
\begin{eqnarray*}
&& \qquad \qquad \ldots \to  H^{1}(M,2\pi\mathbb{Z}) \to H^{1}(M,\mathbb{R}) \xrightarrow{\sigma} \\
&&  H^{1}(M,U(1)) \xrightarrow{\eta}
H^{2}(M,{2\pi}\mathbb{Z})\xrightarrow{\gamma} H^{2}(M,\mathbb{R})
\to \ldots ,
\end{eqnarray*}
where $\textrm{Im}(\eta)=\textrm{Ker}(\gamma)$ is the torsion
subgroup of $H^{2}(M,2\pi\mathbb{Z})$. We recall that classes in
$H^{1}(M,U(1))$ can be identified with the isomorphisms classes of
flat bundles and the torsion classes, but for the zero class,
represent those flat bundles which are not trivial.

\section{The torsion subgroup}
In \cite{minguzzi05b} it has been recognized that  the usual gauge
principle can be generalized in a way compatible with the Standard
Model. In this generalization the transition functions of two
different fields may not necessarily be the same as they could
differ locally by a constant (weak gauge principle). Thus Eqs.
(\ref{q1})-(\ref{q3}) can be suitably generalized.

In a simply connected spacetime this generalization does not lead to
a different physics. Nevertheless, if $H^{2}(M,2\pi\mathbb{Z})$
contains a non-vanishing torsion subgroup then a new interesting
physics arises . Indeed, let us assume that the torsion is a cyclic
subgroup $\langle K \rangle$ of order three (i.e. isomorphic to
$\mathbb{Z}/3$) generated under tensor product by a non-trivial flat
bundle $K$. This subgroup acts on $H^{2}(M,2\pi\mathbb{Z})$
generating orbits containing three elements.

In our formalism the colored components live on vector bundles
associated with different $U(1)$ bundles, the different $U(1)$
bundles being on the orbit generated by the action of $\langle K
\rangle$ on $H^{2}(M,2\pi\mathbb{Z})$. Thus for instance the colors
of the generation $(u,d)$ live in
\begin{eqnarray}
u&:& \quad Q^2 \ \ , \quad Q^2\otimes K \ \ , \quad  Q^2 \otimes K^2 ,\\
d&:& \quad Q^{-1}, \quad Q^{-1}\otimes K, \quad  Q^{-1} \otimes K^2
,
\end{eqnarray}
for a suitable choice of representant $Q$ on the orbit of the $U(1)$
principal bundles associated to $\bar{d}$. The fact that there is a
freedom in choosing the initial principal bundle $Q$, manifests
itself in a cyclic symmetry generated by the replacements $Q\to Q
\otimes K$. The next flavor generations $(c,s)$ and $(t,b)$ live on
the same principal bundles. For instance $c(r)$ is a section on the
same vector bundle that corresponds to $u(r)$, i.e. on a vector
bundle associated to $Q^2$.

To fix the ideas consider the (gauge) transformation of the colored
fields for the flavor $u$ from $U_j$ to  $U_i$
\begin{eqnarray}
A^{i}&=&A^{j}+\dd \beta^{i j}, \label{l1} \\
u^{i}(r) &=& e^{i2\beta^{i j}}\,  u^{j}(r), \label{l2}\\
u^{i}(g) &=& e^{i2\beta^{i j}+i k^{ij} } \,u^{j}(g) , \label{l3}\\
u^{i}(b) &=& e^{i2\beta^{i j}+i 2k^{ij} }\, u^{j}(b) \label{l4}.
\end{eqnarray}
here $e^{ik^{ij}}$ are the constant transition functions of the flat
bundle $K$. Since $K^3$ is a trivial flat bundle $k^{ij} \in
\frac{2\pi}{3}\mathbb{Z}$, up to constant terms $h^i-h^j$ that can
be absorbed with a redefinition of the fields on each set.

\section{The $SU(3)$ symmetry}

As explained in \cite{minguzzi05b}, according to the weak gauge
principle the principal bundle associated to a matter field is not
always the power of a  universal root bundle $Q$. There can be a
flat bundle factor that may differ from matter field to matter
field. It happens that in the particular case considered here, where
all these flat bundles are powers of the non-trivial flat bundle
$K$, the matter fields can be described through a {\em usual} gauge
theory at the price of enlarging the dimension of the gauge group.
Indeed, the previous transformation can be regarded as a $U(1)$
transformation (because the bundle $Q$ has transition functions
$e^{i\beta^{i j}}$) times a $SU(3)$ transformation of matrix
\begin{displaymath}
s_{ij}=\begin{pmatrix}
1 \ \ &  0  &  0\\
 0 \ \ & e^{i k^{ij}}  & 0\\
0 \ \ &  0  & e^{i 2 k^{ij}}
\end{pmatrix} ,
\end{displaymath}
where $s_{ij}$ are interpreted as the transition functions of a
non-trivial $SU(3)$ principal bundle $S=K^{0}\oplus K \oplus K^2$,
which is non-trivial because of the non-triviality of $K$. In this
alternative setting the triplet of fields of different colors are
regarded as a unique section of a 3 (complex) dimensional vector
bundle associated to the non-trivial principal bundle over the group
$U(1) \otimes SU(3)$, $Q \otimes S$.

Summarizing, the existence of non-trivial flat bundles (i.e.
non-trivial elements in the image of $\eta$) in the manifold $M$,
allows for a non-standard gauge theory of the $U(1)$ sector in which
no universal root bundle exists. In order to reestablish the usual
gauge theory one is forced (if possible) to enlarge the
dimensionality of the gauge group so that the matrices $s_{ij}$ can
be interpreted as transition functions of another universal
principal bundle. Thus topological aspects on the behavior of the
$U(1)$ gauge symmetry on the manifold naturally lead to the
introduction of the larger gauge group $U(1)\otimes SU(m)$ where $m$
is the order of the cyclic torsion subgroup generated by $K$ (m=3 in
our case).

One can also keep considering the colored components of a matter
field as distinct sections of vector bundles associated to bundles
of the form $Q^{n}\otimes K^i$. Nevertheless, the interpretation of
the $SU(3)$ symmetry takes in this case a rather unusual form. The
standard treatment regards it as a Lagrangian symmetry under $SU(3)$
transformations of the (vertical) flavor vectors, for instance for
$u$ we have
\begin{displaymath}
\begin{pmatrix}
u'(r) \\
u'(g) \\
u'(b)
\end{pmatrix}=\begin{pmatrix}
U_{rr} & U_{rg} & U_{rb}\\
 U_{gr}& U_{gg}& U_{gb} \\
U_{br} & U_{bg}& U_{bb}
\end{pmatrix}
\begin{pmatrix}
u(r) \\
u(g) \\
u(b)
\end{pmatrix}.
\end{displaymath}
In our case $U$ can not be a matrix whose coefficients are complex
numbers, as $u'(r)$ must live in the same vector bundle as $u(r)$
(and analogously for $u(g)$ and $u(b)$). As a consequence, the
components of $U$ must be taken to be sections of vector bundles
associated to $K^{i}$ with the  following exponents
\begin{displaymath}
\begin{pmatrix}
K^0 & K^2 & K\\
 K& K^0& K^2 \\
K^2 & K& K^0
\end{pmatrix}.
\end{displaymath}
The matrix $U$ becomes a $SU(3)$ matrix only if a local section of
$K$ is chosen. At the intersection $U_{ij}$ the trivializing section
changes with transition function $e^{i k_{ij}}$, and the complex
representant changes as
\begin{displaymath}
\begin{pmatrix}
U_{rr} & U_{rg} & U_{rb}\\
 U_{gr}& U_{gg}& U_{gb} \\
U_{br} & U_{bg}& U_{bb}
\end{pmatrix} \to
\begin{pmatrix}
U_{rr} & U_{rg} e^{i 2 k^{ij}} & U_{rb} e^{i  k^{ij}}\\
 U_{gr}e^{i  k^{ij}}& U_{gg}& U_{gb}e^{i 2 k^{ij}} \\
U_{br} e^{i 2 k^{ij}}& U_{bg}e^{i  k^{ij}}& U_{bb}
\end{pmatrix} .
\end{displaymath}
It is easily seen that a change of section sends $SU(3)$ matrices to
$SU(3)$ matrices. In this model the $U(1)$ vector bundles $K^i$ are
the building blocks from which the $SU(3)$ invariance is
constructed.  We mention that, following a different route, the
authors of \cite{thooft76,lazarides82} also advocated the need for a
unification of the $U(1)$ and $SU(3)$ sectors   in a suitable  gauge
structure. Their strategy differs from ours since we do not need to
enlarge the group $U(1) \otimes SU(3)$ further.

Taking into account that every baryon is a color singlet,  we have
that the principal bundle associated to it has the form $(Q^{3})^k$
for a suitable integer $k$. In particular, it has charge $ke$. For
instance the proton $p=duu'$ has a field
\begin{eqnarray*}
p&=&d(r)u(g) u'(b)+d(b)u(r) u'(g)+d(g)u(b) u'(r)\\
&&-d(r)u(b) u'(g)-d(g)u(r) u'(b)-d(b)u(g) u'(r).
\end{eqnarray*}
The terms in the sum are sections of the same vector bundle
(associated to $Q^3$) and for this reason they can be summed up.
Indeed, the $K$ factors of the $U(1)$ bundles associated with the
different colors cancel in the products. The cancellation is a
general consequence of the fact that the cyclic subgroup $\langle K
\rangle$ is included into $SU(3)$.

The $U(1)$ principal bundles associated with the baryonic fields
have a common root $Q^3$ which is natural to identify with the
$U(1)$ principal bundle root of the principal bundles associated to
the leptonic fields. Therefore, in this model, if quarks are not
taken into account, the usual $U(1)$ gauge theory, with a universal
$U(1)$ bundle $Q^3$ applies.

\section{Initial conditions and dynamics} \label{loj}
Let the spacetime be a direct product between time and space $M=T
\times  S$, $T=\mathbb{R}$. Since $\mathbb{R}$ is contractible the
cohomology groups depend only on the space manifold $S$, for
instance $H^{2}(M,2\pi\mathbb{Z}) \simeq H^{2}(S,2\pi\mathbb{Z})$.
Denote with $\pi_S$ the projection $\pi_S: M \to S$ and let
$\{V_i\}$ be a good covering of $S$. It is convenient to take the
good covering for $M$ defined by $U_i=\pi^{-1}(V_i)$. The previous
isomorphism can be regarded as a consequence of the following

\begin{lemma}
Through suitable changes of sections on each $U_i$ one can make the
functions $\beta^{ij}$ of the gauge transformation (\ref{l1})
between each pair $U_i$, $U_j$, to be independent of time,
$\p_t\beta^{ij}=0$.
\end{lemma}

\begin{proof}
The functions $\beta^{ij}$ on the triple intersections give through
the constant combination $c_{ijk}= \beta^{ij}+\beta^{jk}+\beta^{ki}$
a class $[c_{ijk}]$ on $H^{2}(M,2\pi\mathbb{Z})$ (for the details
see \cite{minguzzi05b}). Differentiating with respect to time
$\p_t\beta^{ij}+\p_t\beta^{jk}+\p_t\beta^{ki}=0$, which means that
$\p_t\beta_{ij}$ are the transition functions of a principal bundle
of fiber $\mathbb{R}$ and group $(\mathbb{R},+)$. Since the fiber is
contractible the bundle is trivial and there are functions
$\p_t\alpha^i$ such that
$\p_t\beta^{ij}+\p_t\alpha^{i}-\p_t\alpha^{j}=0$ or
${\beta'}^{ij}=\beta^{ij}+\alpha^{i}-\alpha^{j}$ is independent of
time ($\alpha^{i}$ can be chosen such that $\alpha^i=0$ at time $t$
so that at the same time ${\beta'}^{ij}=\beta^{ij}$). Thus a change
of section on each $U_i$ with transition function $e^{i\alpha^i}$
makes the functions $\beta^{ij}$ appearing in (\ref{l1}) independent
of time.
\end{proof}

Note that the information of the class to which the matter field,
say $\{ \psi^i\}$, belongs in $H^{2}(M,2\pi\mathbb{Z})$ must be
provided with the initial conditions  $\{ \psi^i\}_{t=0}$, otherwise
the dynamics is in most cases under determined. To see this imagine
for instance an initial condition such that  the component $\psi^i$
is the only one that differs from zero, and the region where it does
not vanish is an open set $A$ such that $A\cap U_j=\emptyset$, for
$j\ne i$. Then the equations of dynamics (suppose they are linear
for simplicity) on $U_i$ will spread out the region on which
$\psi^i$ differs from zero, say $A(t)$, untill $A(t)$ starts having
non empty intersections with $U_j$, $j\ne i$. At this stage in order
to continue the evolution of the field, one has to use the dynamics
equations  on $U_j$ too, and in order to do this, $\psi^i$ must be
transformed into $\psi^j$, a step which requires the knowledge of
the transition functions $e^{iq_{\psi}\beta^{ji}}$ in a given gauge.
In particular we may take the gauge in which $\beta^{ij}$ are
constants.

We conclude that if the dynamics satisfies the gauge principle, i.e.
the Lagrangian is invariant under weak gauge transformations, then
the matter field belongs to a topological sector that must be
specified in the initial conditions and which remains unaltered
during the evolution of the field. Thus the dynamics can not
transform a particle corresponding to a given class of
$H^{2}(S,2\pi\mathbb{Z})$ into a different class. In our application
to the color of quarks this means that the dynamics preserves the
color of a given quark flavor. The class of a particle on
$H^{2}(S,2\pi\mathbb{Z})$ is a true quantum number and in the
interaction of different particles its balance is obtained through
the usual tensor product rule for principal bundles.

In general, given a principal $U(1)$ bundle $P \in
H^{2}(M,2\pi\mathbb{Z})$, the principal bundles on the orbit
generated by the action of $\langle K \rangle$,
\begin{equation}
P, \quad P \otimes K, \quad P \otimes K^2 ,
\end{equation}
represent different topological (vacuum) sectors of the theory. In
particular,  we may consider the hypothesis that the transition
between them requires a huge amount of energy. As a consequence of
this assumption all the baryons, sharing a common square root $Q^3$
would live in the same topological vacuum state, while the isolated
quarks would not. The interaction between baryons would follow the
usual rules of the Standard Model but the free quarks would not be
seen as the energy required to move the state from one topological
sector to the other would be too high. Examples of this topological
mechanism are familiar in physics, they run from the simple
kink-antikink example to the physics of instantons
\cite{rajaraman82}.

\section{The topology of the Universe} The present theory works
only in a spacetime manifold such that $H^{2}(M,2\pi\mathbb{Z})$
contains  a cyclic torsion subgroup of order three. This fact could
provide a constraint for current investigations
\cite{reboucas04,aurich05} on the space topology of the Universe
under the assumption $M=S \times \mathbb{R}$. Unfortunately, the
recently suggested \cite{luminet03,aurich05b} Poincar\'e
dodecahedral topology, $S=\mathcal{P}=S^3/\Gamma_{\mathcal{P}}$ does
not share this property since $H^{2}(\mathcal{P},2\pi\mathbb{Z})=0$.

The lens space $S=L(3,p)$, for $p=1,2$ (L(3,2) is the `mirror image'
of $L(3,1)$) is obtained from $SU(2) \sim S^3$ by making the
quotient with respect to the action of
\begin{displaymath}
\begin{pmatrix}
\omega & 0 \\
& \omega^{p}
\end{pmatrix} ,
\end{displaymath}
with $\omega\in \mathbb{C}$, $\omega^{3}=1$. It provides an example
of a homogeneous space compatible with the required cohomology.
Indeed, using results from \cite{munkres84}, we find
\begin{displaymath}
H^{j}(L(3,p),2\pi\mathbb{Z})\simeq \left\{\begin{array}{ll}
2\pi\mathbb{Z}, & \textrm{if} \ j=0 \ \textrm{or} \ j=3,\\
2\pi\mathbb{Z}/3,  & \textrm{if} \ j=2, \\
0,& \textrm{otherwise}.
\end{array} \right.
\end{displaymath}
If the topological scale of the universe, represented by the
injectivity radius $r_{inj}$,  is smaller that the radius of the
surface of last scattering $\chi_{LSS}$, then the non-trivial
topology of the universe would be detectable using the CMB-circles
in the sky method \cite{cornish98}. Both length scales have been
studied in deep for many candidate topologies \cite{mota03b}. In the
case of the lens space $L(3,1)$ it has been found that the
injectivity radius is larger than the radius of the last scattering
surface as derived from  latest cosmological data. Indeed, $L(3,1)$
is a single action manifold \cite{gausmann01} with a cyclic holonomy
group of order $n=3$ which implies \cite{weeks03},  $r_{inj}=\pi/n$.
In \cite{weeks03} it has been shown that in this case $\chi_{LSS}
>r_{inj}$, only if $\Omega_{tot} \gtrsim 1+1/n^2$. Present estimates of
$\Omega_{tot}$ give $\Omega_{tot}=1.02 \pm 0.02$, which rules out
the possibility of detecting the cosmic topology in the $L(3,1)$
case. The last search of pairs of matching circles in the sky has
given a null result \cite{cornish04} and is therefore compatible
with a lens shaped Universe $L(3,1)$.

\section{Conclusions}
In a topologically non-trivial manifold it is possible to  weaken
the gauge principle allowing for $U(1)$ gauge transformations that
only coincide locally for each field up to a constant
field-dependent phase. In this way the flat bundles over the
manifold become relevant and two cohomology classes on
$H^{2}(M,2\pi\mathbb{Z})$, i.e. two particles, may  differ  by the
tensor product  with a (flat) torsion class. Here we have
investigated the example of a torsion subgroup of order three
identifying the orbits of the torsion subgroup with the colors of
the same quark flavor.


The theory, as much as baryons and leptons are concerned, coincides
with the usual Standard Model gauge theory. Indeed, we have shown
that the baryons live on the same topological background $Q^3$ while
the free quarks  live in different topological backgrounds. We have
argued that a high transition energy between those backgrounds would
give a natural mechanism for the quark confinement. The theory
naturally embodies two charge quantization units which can be
identified with (i) (minus) the electron  charge $e$, which turns
our to be the fundamental charge for those particles that live in
the same background (baryons and leptons), and (ii) the electric
quantization unit $e/3$.

The theory makes hypothesis on the torsion subgroup of
$H^{2}(M,2\pi\mathbb{Z})$ and hence on the topology of the Universe.
As a working assumption we assumed that the non-trivial topology
manifests itself at the cosmological scale. Thus we tested the
simplest homogeneous space which satisfies our topological
constraint, i.e. the lens space $L(3,1)$. Although according to some
authors it is not favored \cite{luminet03}, this space has not yet
been ruled out by the latest observations \cite{cornish04}. In any
case the lens space $L(3,1)$ represents only the simplest
possibility compatible with this theory. More work is needed to
extend our study to more general manifolds which satisfy the
topological constraint.

\section*{Acknowledgments}
Very useful comments by a referee are acknowledged.

%

\end{document}